\documentclass[aps,prd,10pt,twocolumn,showpacs]{revtex4-1}
\usepackage{graphics}  
\usepackage{graphicx}  
\usepackage{epsfig}

\begin{document}
\clearpage

\title{High-energy neutrino fluxes and flavor ratio in the Earth's atmosphere}

\author{T.S. Sinegovskaya\,${}^{1}$, A.D. Morozova\,${}^2$} 
\author{S.I. Sinegovsky\,${}^3$}
\email[E-mail: ]{sinegovsky@api.isu.ru}
\affiliation{$^1$ Irkutsk State Railway University, 664074 Irkutsk, Russia \\ 
$^2$ Physics Faculty, Irkutsk State University, 664003 Irkutsk, Russia  \\ 
$^3$  Institute of Applied Physics, Irkutsk State University, 664003 Irkutsk, Russia}

\begin{abstract}

We calculate the atmospheric neutrino fluxes in the energy range $100$ GeV -- $10$ PeV with the use of several known hadronic models and a few parametrizations of the cosmic ray spectra which take into account the knee. The calculations  are  compared with the atmospheric neutrino  measurements  by  Frejus, AMANDA, IceCube and ANTARES. An analytic description is presented for the conventional ($\nu_\mu+\bar\nu_\mu$) and ($\nu_e+\bar\nu_e$) energy spectra, averaged over zenith angles, which can be used  to obtain test data of the neutrino event reconstruction in neutrino telescopes.  
The sum of the calculated atmospheric $\nu_\mu$ flux and the IceCube best-fit astrophysical flux  gives the evidently higher flux as compared to the IceCube59 data, giving rise the question concerning the hypothesis of the equal flavor composition of the high-energy astrophysical neutrino flux. 
Calculations show that the transition from the atmospheric electron neutrino flux to the  predominance of the astrophysical neutrinos occurs at $30-100$ TeV if the prompt neutrino component is taken into 
consideration. The neutrino flavor ratio, extracted from the IceCube data, does not reveal the trend to increase with the energy as is expected for the conventional neutrino flux in the energy range $100$ GeV -- $30$ TeV. A depression of the ratio $R_{\nu_\mu/\nu_e}$ possibly indicates that the atmospheric electron neutrino flux obtained in the IceCube experiment contains an admixture of the  astrophysical neutrinos in the range $10-50$ TeV.


\end{abstract}
\maketitle                          

\pagenumbering{arabic}
\setcounter{page}{1}
\maketitle

\section{Introduction}

High-energy neutrinos produced in decays of pions, kaons, and charmed particles
of the extensive air shower induced by cosmic rays passing through the Earth atmosphere,  form an unavoidable  background for the detection of astrophysical neutrinos. Sources of extra-terrestrial high-energy neutrinos is the challenge to solve to solve which large-scale neutrino telescopes, NT200+~\citep{nt200-08}, IceCube~\cite{icecube11, icecube40_lim, IC_dif2011},  ANTARES~\cite{antares13a, antares13b} are designed. 
The high-energy atmospheric neutrinos became accessible  to the  experimental studies only last years.
By now, the energy spectrum of high-energy atmospheric muon neutrinos has been measured  in  the four experiments: Frejus~\cite{frejus} at energies up to 1 TeV, AMANDA-II~\cite{amanda10} in the energy range 
 $1-100$ TeV, IceCube~\cite{icecube11,IC59_numu} at $100$ GeV -- $575$ TeV, and ANTARES at energies  $100$ GeV -- $200$ TeV ~\cite{antares13a}. 
Not long ago the IceCube presented also results for the electron neutrino spectrum  measured in the energy range $\sim 80$ GeV -- $6$ TeV~\cite{ice79_nue}. 
Thus a possibility appeared to evaluate the neutrino flavor ratio using the IceCube measurements and to compare this one with  predictions. 

Lately IceCube found the $37$ high-energy neutrino events~\cite{pevnu2013, science342, 3years} in energy range $30$ TeV - $2$ PeV,  most of which are hardly consistent with events expected from the atmospheric muons and neutrinos, $8.4\pm 4.2$ and $6.6^{+5.9}_{-1.6}$  respectively. 
The  neutrino events, three of which have energies above $1$ PeV  were detected over the three years  2010-2013 ($988$ days), give the statistical significance of their astrophysical origin at the level of  $5.7 \sigma$. 
 
After the IceCube reported~\cite{pevnu2013} on the detection of the first two neutrino-induced events  
with deposited energy 1.04 and 1.14 PeV, the prompt analysis of the origin of the highest energy neutrinos ever detected was performed~\cite{murase13a, murase13b, stecker13, kalash13, halzen14a, halzen14b, tjus14, winter14}. 

Increasing with energy contribution of charmed particles to the atmospheric neutrino flux becomes the  source of the large uncertainty at energies above $\sim 200$ TeV for muon neutrinos, and above $
\sim 20$ TeV for electron neutrinos. More complicated picture  is likely if the astrophysical neutrinos and the atmospheric conventional/prompt ones are entangled. 
It is quite possible that astrophysical neutrino flux becomes dominant over  the atmospheric  electron neutrino flux at energies  $20-50$ TeV.
Thus a comparison of atmospheric neutrino spectra  calculated for various hadron-interaction models with high-energy neutrino spectra measurements can shed light on the most uncertain constituent of the atmospheric neutrino background, in spite of large statistical and systematic uncertainties.

Here we calculate atmospheric neutrino fluxes at energies $10^2-10^7$ GeV for zenith angles from $0^\circ$ to $90^\circ$ as well as the angle averaged  spectrum with the use of high-energy hadronic interaction models QGSJET~II-03~\cite{qgsjet2} and  SIBYLL 2.1~\cite{sibyll}.  These models are widely employed to simulate extensive air showers (EAS) with the Monte Carlo method, and  were also applied to compute the cosmic-ray hadron and muon fluxes~\cite{KSS08, kss13}. Besides, in this work computation we  employ also the old known hadronic model by Kimel \& Mokhov (KM)~\cite{KMN} which was checked by comparison of the calculated atmospheric  hadron and muon spectra with the experiment~\cite{KSS08}.  

The calculation has been performed for three parametrizations of the measured spectra and composition of primary cosmic rays in the  energy range comprising the knee: 1)  the model by Zatsepin \& Sokolskaya (ZS)~\cite{ZS3C}, 2) the model by Hillas~\cite{Hillas06} and Gaisser (HGm)~\cite{HiG2012}, and  3) the modified multi-knee (polygonato) model (BK)~\cite{polygon, BK2011}.
Below we denote, for short,  different versions of the calculations  according to the pattern ``a primary cosmic ray spectrum + an hadronic interaction model'' or vice versa, ``an hadronic interaction model + a primary cosmic ray spectrum''. For example:  HGm+QGSJET (or HGm+QGS2), ZS+QGS2, HGm+KM, SIBYLL+BK, and so on. 

The paper is organized as follows.
In Sec.~II  we give a sketch of the computation scheme:  a short review  of the calculation method and
the data input, primary cosmic-ray spectra and hadronic interaction models used in the computation. 
In Sec.~III presented are the results of calculation of the atmospheric muon neutrino energy spectrum  and comparison with experiments. In Sec.~IV  we compare calculations of the electron neutrino spectrum and the neutrino flavor ratio $R_{\nu_\mu/\nu_e}$ with the IceCube measurements.  Main results of the paper are discussed in  Sec.~V.

\section{Scheme of the computation} 

\subsection{Outline of the calculation method}

To compute the atmospheric muon and electron neutrino spectra  we employ 
the method to solve the hadronic cascade equations~\cite{ns2000, KSS08} which makes it possible to consider the non-power behavior of the cosmic-ray spectrum, energy dependent
total inelastic cross sections $\sigma_{hA}^{in}(E)$ of hadron-–nucleus interactions, and 
 a violation of the Feynman scaling law for the inclusive cross sections.  
 The calculation procedure was validated through the careful comparison with experiments  of the calculated atmospheric hadron fluxes and the sea-level muon spectrum which covers the wide range of energies for different zenith angles ~\cite{KSS08, kss13}. 
Omitting details we give here  a brief review of the calculation scheme  aimed to introduce generalized $Z$-factors.  

The differential energy spectra of the  secondary protons $p(E,h)$ and neutrons $n(E,h)$ at atmospheric
depth $h$ obey the equations 
\begin{eqnarray}\nonumber
\frac{\partial N^{\pm}(E,h)}{\partial h}&=&-\frac{N^{\pm}(E,h)}{\lambda_N(E)} + \frac{1}{\lambda_N(E)} \\
  &\times & \int_0^1\Phi^{\pm}_{NN}(E,x)N^{\pm}(E/x,h)\frac{dx}{x^2},\label{nucleons}
\end{eqnarray}
where $N^{\pm}(E,h)=p(E,h)\pm n(E,h)$,
\[ 
\Phi^\pm_{NN}(E,x)=\frac{E}{\sigma_{pA}^{in}(E)}\left[\frac{d\sigma_{pp}(E_0,E)}{dE}\pm
\frac{d\sigma_{pn}(E_0,E)}{dE}\right], 
\]
 $\lambda_N(E)=\left[N_0\sigma_{pA}^{in}(E)\right]^{-1}$  is  the  nucleon  interaction length,  $N_0=N_A/A$ is the number of nuclei per gram of air, $x=E/E_0$  is the fraction of the primary nucleon energy $E_0$  carried away by the secondary nucleon,
$d\sigma_{ab}/dE$ is a cross section for the inclusive reaction $a+A\rightarrow b+X$, integrated over the transverse momentum. The boundary conditions for Eq.~(\ref{nucleons}) are:  $N^\pm(E,0)=p_0(E)\pm n_0(E)$. 

Let us seek a solution of system~(\ref{nucleons}) in the form 
\begin{equation}\label{ans}
N^\pm(E,h)=N^\pm(E,0)\exp\left[ -\frac{h(1-{\cal Z}^\pm_{NN}(E,h))}{\lambda_N(E)}\right],	
\end{equation}
where  ${\cal Z}^\pm_{NN} (E,h)$ are unknown functions.
Substitution of Eq.~(\ref{ans}) into~(\ref{nucleons}) yields equations for the functions
${\cal Z}^\pm_{NN}$ (${\cal Z}$-factors):
	\begin{eqnarray}\label{difzf}\nonumber
	\frac{\partial(h{\cal Z}^\pm_{NN})}{\partial h}&=&\int_0^1 dx\,\Phi^\pm_{NN}(E,x)\eta^\pm_{NN}(E,x) \\
	 &\times&	\exp\left[-h{\cal D}^\pm_{NN}(E,x,h)\right],
		\end{eqnarray}
where $ \eta^\pm_{NN}(E,x)=x^{-2}N^\pm(E/x,0)/N^\pm(E,0),$
		\begin{equation}\label{Dpm}
		 {\cal D}^\pm_{NN}(E,x,h)=\frac{1-{\cal Z}^\pm_{NN}(E/x,h)}{\lambda_N(E/x)}-
		\frac{1-{\cal Z}^\pm_{NN}(E,h)}{\lambda_N(E)}.
		\end{equation}	
Integrating Eq.~(\ref{difzf}), one obtains a nonlinear integral
equation,		
 \begin{eqnarray}\label{zfactor}\nonumber
		{\cal Z}^\pm_{NN}(E,h)&=&\frac{1}{h}\int_0^hdt\int_0^1 dx\,\Phi^\pm_{NN}(E,x)\eta^\pm_{NN}(E,x) \\
	&\times	&\exp\left[-t{\cal D}^\pm_{NN}(E,x,t)\right],
		\end{eqnarray}
which  we  solve using the iterative approach. 
Choosing as the starting point  
 ${\cal Z}^{\pm(0)}_{NN}(E,h)=0$, i.e.
		${\cal D}^{\pm(0)}_{NN}(E,x,h)=1/\lambda_N(E/x)-1/\lambda_N(E),$
we find for the $n$th step 
\begin{eqnarray}\nonumber
{\cal Z}^{\pm(n)}_{NN}(E,h) &=& \frac{1}{h}\int_0^hdt\int_0^1 dx \Phi^\pm_{NN}(E,x)\eta^{\pm}_{NN}(E,x) \\
	&\times & 	\exp\left[-t{\cal D}^{\pm(n-1)}_{NN}(E,x,t)\right],	
\end{eqnarray}
\begin{eqnarray} \label{znn-5}\nonumber
		{\cal Z}^{\pm(n)}_{NN}(E,h)&=&\frac{1}{h}\int_0^hdt\int_0^1 dx \Phi^\pm_{NN}(E,x)\eta^{\pm}_{NN}(E,x) \\
	&\times &	\exp\left[-t{\cal D}^{\pm(n-1)}_{NN}(E,x,t)\right],
		\end{eqnarray}				
where	
		\begin{eqnarray}\nonumber
{\cal D}^{\pm(n-1)}_{NN}(E,x,h)&=&\frac{1-{\cal Z}^{\pm(n-1)}_{NN}(E/x,h)}{\lambda_N(E/x)} \\
&-&\frac{1-{\cal Z}^{\pm(n-1)}_{NN}(E,h)}{\lambda_N(E)}.
	\end{eqnarray}	

The functions  ${\cal Z}(E,h)$ depend on two variables unlike the commonly known  $z$-factor and carry   imprints of cosmic-ray spectra, hadron-nuclei interactions  and the hadronic cascade evolution in the atmosphere. In the case of the power-law cosmic-ray spectrum, the Feynman scaling  for the  hadron production cross sections and $\sigma_{hA}^{in}=$ const, the function ${\cal Z}_{ab}(E,h)$ is reduced to a constant or $z_{ab}(E)$ (see details in Refs.~\cite{kss13, ijmpa2010}).%

In similar fashion  meson cascade equations can be solved using the nucleon and meson sources~\cite{KSS08}.
The main pion sources in the atmospheric shower are the interactions of nucleons and pions with
air nuclei and kaon decays. As a first step, the meson component can be detached from the nucleon one by
neglecting the small contribution from the $N\bar N$ pair production in meson--nucleus collisions. 
At this step, the pion equations are detached from the kaon ones  with neglect of kaon decays to pions as well as of the pion production by kaons. 
In turn, the kaon part of cascade contains the nucleon source as the main one and pion sources as additional source in as much as the reaction $\pi+A \rightarrow K+X$ is taken into account.  

At the second step,  one can calculate small corrections to the nucleon,
pion and kaon fluxes, allowing for (i)  additional pion source from decays of kàons to pions,
(ii) pion production in kaon-nuclei collisions, $K+A \rightarrow \pi+X$, and (iii) $N\bar N$ pair production. 

The Genz–-Malik adaptive cubature algorithm~\cite{genz-malik} turned out to be very useful in a numerical 
realization of the method for multidimensional integration; a fast algorithm based on quadratic 
B-splines was used to interpolate and approximate the calculated functions. Note the high convergence 
of the method: in the nucleon component calculations,  five  iterations   are  required   to achieve an accuracy close to  $1\%$, while two iterations are enough to achieve the same accuracy in the meson flux computations.
 
\subsection{Primary cosmic-ray spectra}

As the primary cosmic ray spectra and composition in wide energy range we use in our calculations following models: (1)  the model by Zatsepin \& Sokolskaya (ZS), (2) the novel cosmic-ray (CR) spectrum approximation (HGm) by Hillas ~\cite{Hillas06} and  Gaisser~\cite{HiG2012},  and  (3) the modified multi-knee model  by Bindig, Bleve and Kampert (BK)~\cite{BK2011}  based on KASCADE  data~\cite{KASCADE05} and  the  polygonato model by   H$\ddot{\text o}$randel~\cite{polygon}.

The Zatsepin and Sokolskaya model~\cite{ZS3C} comprises contributions to the cosmic ray flux of three classes of Galaxy astrophysical sources: 
(I) isolated (nonassociated) supernovae (SNe) exploding into random interstellar medium (ISM),  the magnetic rigidity $R = ECR /eZ < 50$ TV;
(II) the  most powerful sources of CR, which are high mass SNe exploding into ISM in OB associations that give rise to particles with energies up to $4Z$ PeV:  CR particles are accelerated by shock waves passing through the stellar wind, i. e. OB star explodes into a dense ISM;
(III)  weak sources owing to explosions of novae,  $R < 200$ GV. 
 
All of three classes of CR sources produce power-law energy spectra with different spectral indices  $\gamma=\alpha +0.33$, where $\alpha$ is the index of source spectrum. In the region of effective acceleration, spectral indices are  $2.63$ (class I), $2.43$ (II), and  $2.90$ (III). 
In the energies range $E=100$ TeV - $1$ PeV, fluxes of  p and He rise ($\gamma_{\rm{p,He}} < 2.75$),
median nuclei (CNO) and heavy ones fall ($\gamma_{\rm{CNO}} > 2.75$).
At the energies above the CR ``knee'' ($E > 3$ PeV) p and He diminish, and heavy nuclei grow.  Fe nuclei dominate at  $E > 30$ PeV.

The model ZS describes well data of the ATIC2 direct measurements~\cite{atic2_07} in the range  $10-10^5$  GeV and gives a motivated extrapolation of these data up to  $100$ PeV  -- the energy region where the cosmic ray spectra and elemental composition are derived from  measured characteristics of EAS. 
The ZS proton and helium spectra at $E\gtrsim 10^6$ GeV are compatible with  KASCADE  spectra, reconstructed with the usage of hadronic models QGSJET01 and SIBYLL.  ZS spectra well agree also with the Hillas-Gaisser model up to $1$ PeV. 

Since direct  measurements  of the cosmic ray spectra and elemental composition are terminated close to
$~100$ TeV, one needs to  make the spectrum extrapolation to high energies, above the knee.
The model  by Hillas and  Gaisser~\cite{Hillas06, HiG2012} includes three classes of sources:
 i) supernova remnants in the Galaxy, ii)  Galaxy high-energy sources of still uncertain origin which contribute to the cosmic ray flux between the ``knee'' ($3$ PeV) and the  ``ankle'' ($4$ EeV), 
 iii) extragalactic astrophysical objects  (Active Galactic Niclei, sources of the gamma-ray bursts and others).   
 
The composite spectrum is formed of five groups of nuclei (p, He, CNO, Mg-Si and Mn-Fe).
Each of the three populations accelerates five groups of nuclei, the spectrum of which cuts off at a characteristic rigidity.
 The parameters for the class  $1$ spectrum were taken  from  CREAM measurements~\cite{cream10}
and  extrapolated (to a rigidity of $4$ PV) to take into account the  knee. 
The extrapolation is consistent with measurements of the all-particle spectrum  beyond the knee in the EAS experiments. The extragalactic component takes into  account also the  measurement data by HiRes, PAO and Telescope Array.  
 In our calculations, we use the version with mixed composition for extragalactic sources of cosmic rays, denoted here as HGm,  which corresponds to the H3a of Ref.~\cite{ice59_lim}).
More details concerning this parametrization one can find also in Ref.~\cite{fedyn12}. 
 
The polygonato model~\cite{polygon} comprising only galactic sources  is the third known cosmic-ray model used in our calculations. 
The modified multi-knee (polygonato) model  by Bindig, Bleve and Kampert~\cite{BK2011} was aimed to take into account the KASCADE  data~\cite{KASCADE05} concerning elemental composition  of cosmic rays  around the knee.

\subsection{High-energy hadronic interaction models}

Calculations of hadronic cascades induced by high-energy cosmic-ray particles and the atmospheric neutrino flux of the PeV scale require either an extrapolating cross sections measured at lower energies or developing the models to give the reliable predictions at high and ultrahigh energies. 
Direct measurements of inclusive cross sections for the nucleon and meson production in hadron-nucleus collisions are still far from being complete because of the restricted  kinematics region of LHC cross section measurements. 

In this work, we apply known hadronic interaction models QGSJET II-03~\cite{qgsjet2}  and SIBYLL 2.1~\cite{sibyll}, which currently undergo overall test in two ways:  (i) through simulations of EAS induced by high-energy cosmic-ray particles and  (ii) through comparison of the model predictions with results of LHC experiments.  

Besides,  we also use the hadronic interaction model proposed by Kimel and Mokhov  (KM)~\cite{KMN}, for which we adopt updated parameters~\cite{ns2000, Naumov2001, FNV01}. Based on accelerator data at energies up to $1.5$ TeV, KM however obeys the Feynman scaling law and can be valid for higher energies, making predictions compatible with QGSJET and  SIBYLL.
The predictions of KM model for $pp$  and $pA$  interactions are also close to results obtained  in
hadronic model DPMJET-III~\cite{mc_code00, dpmjet07} based (like SIBYLL) on the dual parton model. 
The KM model was also applied in three-dimensional Monte Carlo calculations of the atmospheric neutrino fluxes~\cite{Derome}.  
\begin{table}[!b]
\caption{\label{tab_z} $z_{pc}$-factors for $\pi$ and $K$ mesons.}
\begin{ruledtabular}
  \begin{tabular}{lccccc}
   Model  & $ E_0 $, GeV & $ z_{p\pi^+} $  & $ z_{p\pi^-} $  & $ z_{pK^+} $  & $ z_{pK^-} $ \\
   \hline 
          & $10^2$    & 0.043     & 0.035  & 0.0036   &  0.0030    \\
 QGSJET   & $10^3$    & 0.036     & 0.029  & 0.0036   &  0.0028   \\
  II-03   & $10^4$    & 0.033     & 0.028  & 0.0034   &  0.0027    \\
            \hline
          & $10^2$    & 0.036     & 0.026  & 0.0134   &  0.0014     \\
 SIBYLL   & $10^3$    & 0.038     & 0.029  & 0.0120   &  0.0023     \\ 
  2.1     & $10^4$    & 0.037     & 0.029  & 0.0097   &  0.0027     \\ 
            \hline  
          & $10^2$    & 0.044     & 0.027  & 0.0051  &  0.0015   \\
   KM     & $10^3$    & 0.046     & 0.028  & 0.0052  &  0.0015   \\ 
          & $10^4$    & 0.046     & 0.029  & 0.0052  &  0.0015   \\   
            \hline 
 DPMJET   & $10^3$   &$0.04 $ & $ 0.035$  & $ 0.0070$   & $ 0.0035$    \\
 III      & $10^4$   &$0.04  $   & $  0.035$  & $ 0.0070$ & $ 0.0031$   \\   
 \end{tabular}
 \end{ruledtabular}
  \end{table} 

 The current models differ in accuracy of description of soft interactions and contribution of the  diffraction dissociation, in ways to account for semihard interactions (minijets), in description of hadron-nucleus interactions (Glauber approach or Glauber-Gribov theory, the superposition or semisuperposition picture), in the degree of the scaling violation at high energies. 

To illustrate in part the difference of the hadronic models one can compute  the cosmic ray  spectrum-weighted moments ($z$-factors) for proton-air interactions $pA\rightarrow cX$ of the inclusive spectra
 $(x/\sigma^{in}_{pA})\,{(d\sigma_{pc}}/{dx})$:
\begin{equation}\label{mom}
z_{pc}(E_0)
=\int\limits_0^1\frac {x^{\gamma}}{\sigma^{in}_{pA}}\frac {d\sigma_{pc}}{dx}\,dx,
\end{equation}
where  $x=E_c/E_0$, $\gamma=1.7$,  $c=\pi^\pm, K^\pm$; 
z-factors for DPMJET-III are borrowed from Ref.~\cite{dpmjet07}.
As one can see from Table~\ref{tab_z}, $z$-factors obey the approximate scaling law  in  DPMJET-III, KM and SIBYLL 2.1 (except the production of $K$-mesons), while in case of QGSJET-II noticeable violation of the scaling is found just for pions. 
%

\section{Fluxes of atmospheric muon neutrinos \label{sec:numu}} 

The calculation is performed on the basis of the method~\cite{ns2000, KSS08} of solution of the hadronic cascade equations in the atmosphere taking into account the non-power energy spectrum of the cosmic rays, a violation of Feynman scaling and a growth with the energy of the total inelastic cross sections of hadron-nucleus collisions.

Along with major sources of the muon neutrinos, the $\pi_{\mu2}$ and  $K_{\mu2}$ decays, we consider three-particle semileptonic decays of charged and neutral kaons, $K^{\pm}_{\mu3}$ (the branching ratio $3.32\%$), $K^{0}_{\mu3}$ ($27\%$). 
Moreover we account for small contributions originating from decay chains
$K\rightarrow\pi\rightarrow\nu_\mu$ ($K^0_S\rightarrow \pi^+\pi^-$, $K^\pm \rightarrow \pi^\pm \pi ^0$), 
and from the muon decays. 
\begin{figure}[!b]
\includegraphics[width=80 mm]{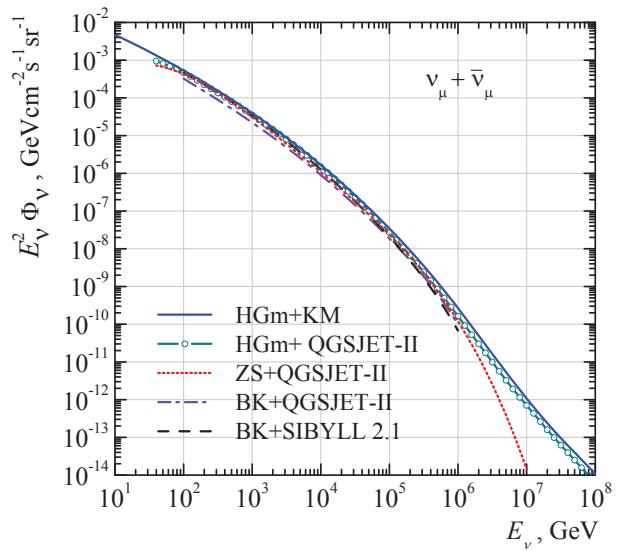} 
\caption{\label{compar_3pcr}Comparison of atmospheric $\nu_{\mu}+\bar\nu_{\mu}$ fluxes calculated  using three hadronic models, KM, QGSJET-II-03, SIBYLL 2.1, and  three  parametrizations of the cosmic-ray spectrum.}
\end{figure}
\begin{figure*}[!t]
\includegraphics[width=85 mm]{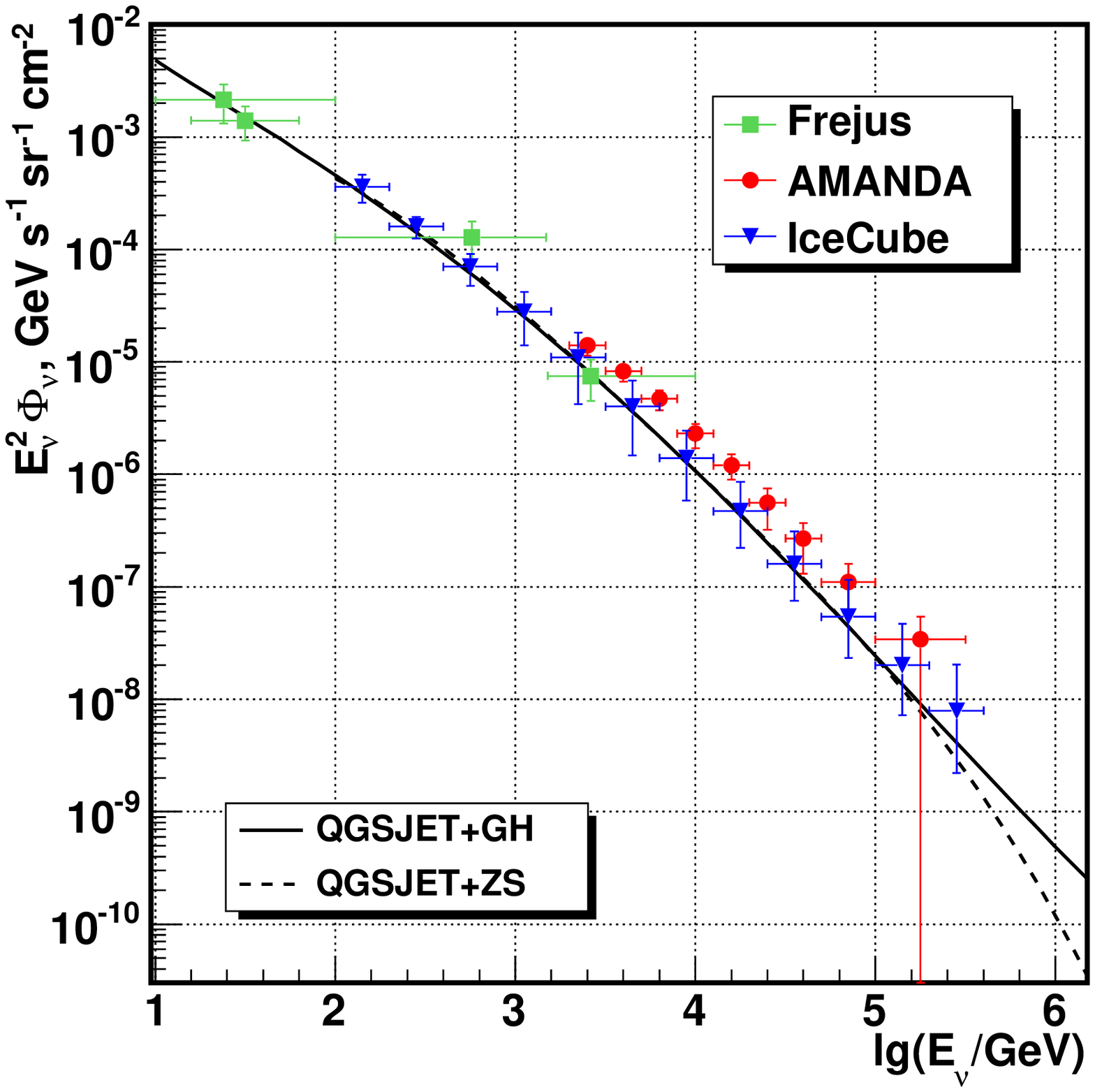} 
\includegraphics[width=85 mm]{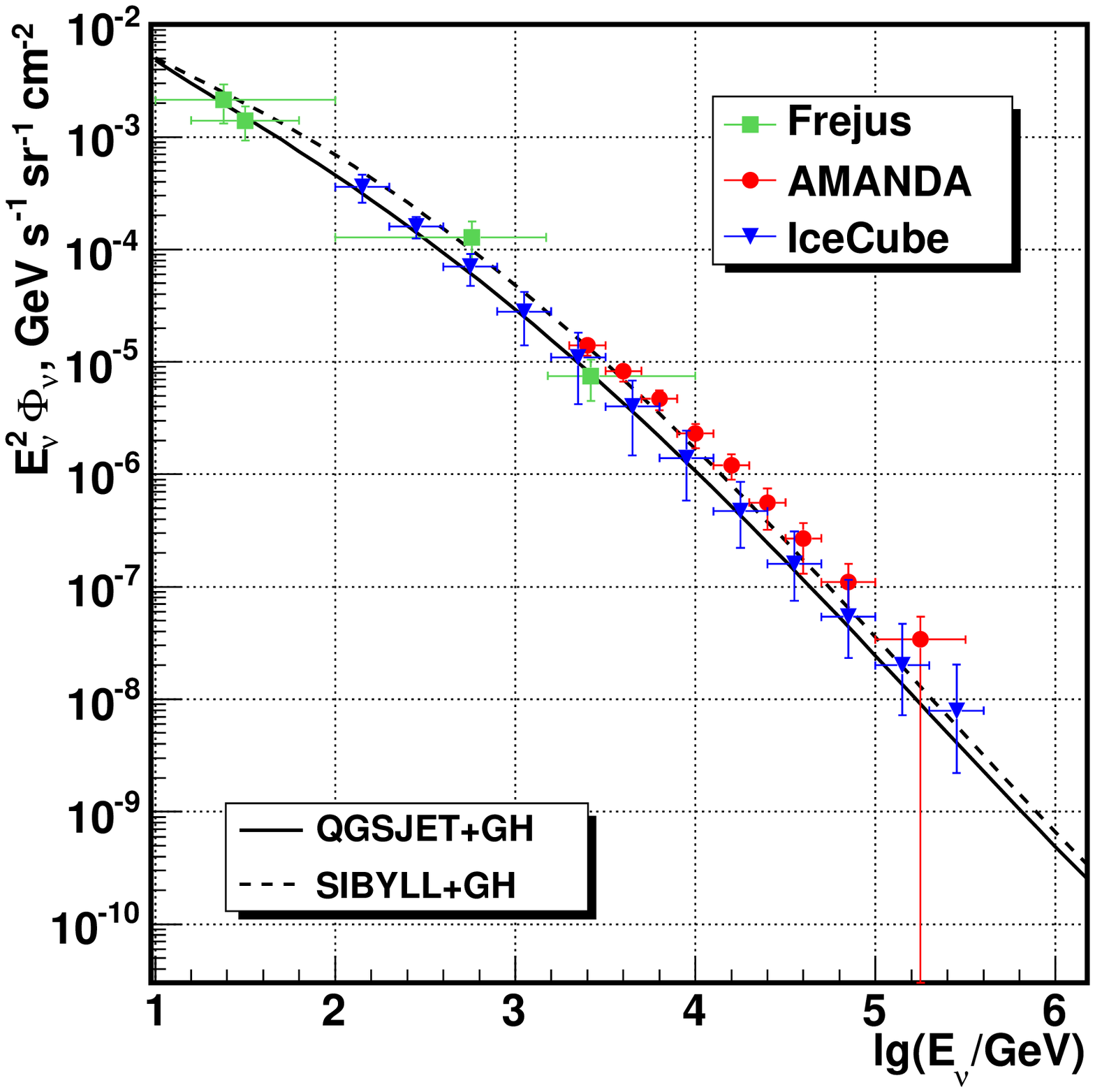}
\caption{Conventional $\nu_\mu+\bar\nu_\mu$ spectrum averaged over zenith angles: an influence of the  cosmic ray spectrum ``knee'' (left panel) and the hadronic model (right). 
Symbols: data of Frejus~\cite{frejus}, AMANDA-II~\cite{amanda10} and  IceCube~\cite{icecube11} experiments.}
\label{qgs_2pcr}       
\end{figure*}
\begin{figure*}[!t]  
\includegraphics[width=80 mm]{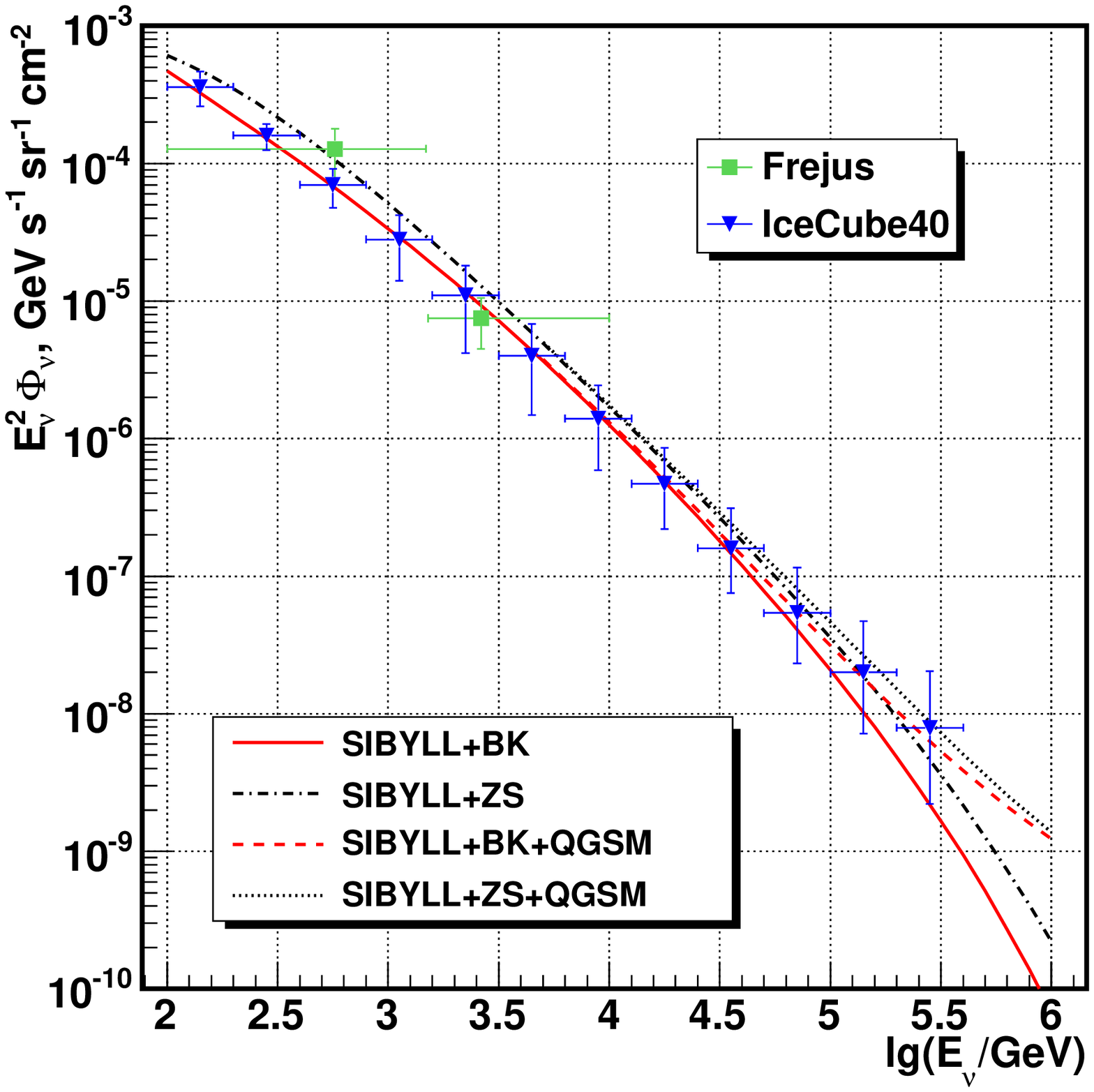} \hskip 0.5 cm  
\includegraphics[width=85 mm]{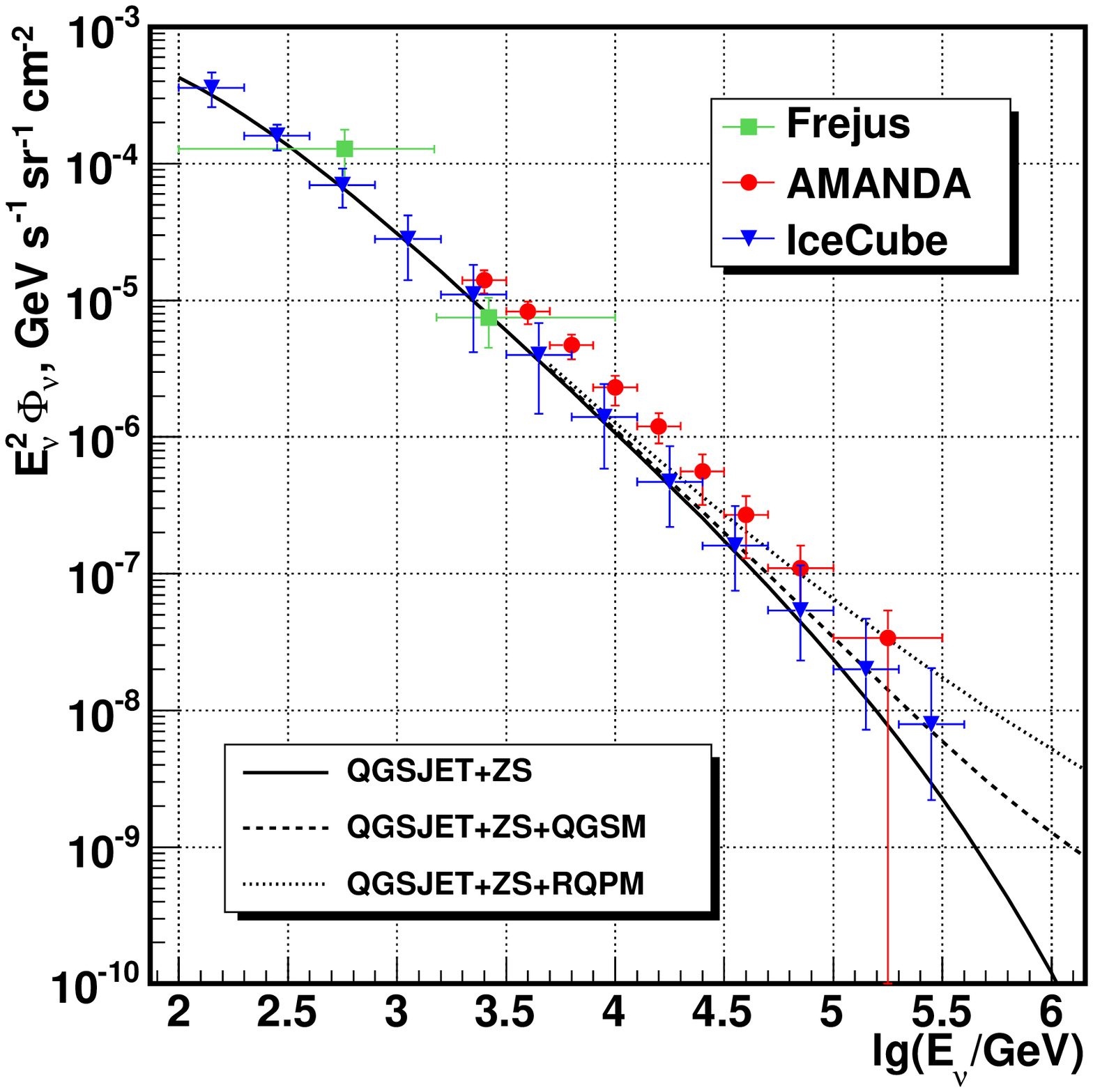}
 \caption{ \label{numu_sib} Fluxes of the conventional and prompt (QGSM, RQPM) muon neutrinos ($\nu_\mu + \bar{\nu}_\mu$) calculated  with usage of  SIBYLL 2.1 (left panel)) and QGSJET~II (right panel))  for the cosmic-ray spectrum parametrizations involving the knee.}          
 \end{figure*}
 
The comparison of the muon neutrino fluxes calculated with three recent primary spectrum models  (Fig.~\ref{compar_3pcr}) shows that they are rather close each other up to $1$ PeV. 
 \begin{table}[!t] 
  \caption{\label{tab_flux} The neutrino flux ratio  calculated with SIBYLL 2.1, QGSJET~II-03 and KM for cosmic ray spectra ZS~\cite{ZS3C} and HGm~\cite{HiG2012}: sib/qgs2 -- column 1 (4); km/qgs2 -- 2 (5); 
  sib/km -- 3 (6).} 
    \begin{ruledtabular}
  \begin{tabular}{c|ccc|ccc} 
     $E_{\nu}$, GeV  & $ 1 $ & $ 2 $ & $ 3 $ & $ 4 $ & $ 5 $ & $ 6 $ \\\hline
 &\multicolumn{3}{c|} {ZS: ($\nu_\mu+\bar\nu_\mu$)} & \multicolumn{3}{c} {ZS:  ($\nu_e+\bar\nu_e$)}\\ 
    $10^3$  & 1.70 & 1.05 & 1.62 & 1.48  & 0.85 & 1.74 \\
    $10^4$  & 1.53 & 1.04 & 1.47 & 1.39  & 0.81 & 1.72 \\
    $10^5$  & 1.53 & 1.10 & 1.39 & 1.35  & 0.95 & 1.42 \\
    $10^6$  & 1.79 & 1.64 & 1.09 & 1.48  & 1.56 & 0.96 \\
    $10^7$  & 1.85 & 2.08 & 0.89 & 1.45  & 1.91 & 0.76 \\ \hline
 &\multicolumn{3}{c|} {HGm: ($\nu_\mu+\bar\nu_\mu$)} & \multicolumn{3}{c}{HGm: ($\nu_e+\bar\nu_e$)}\\ 
    $10^3$  & 1.59 & 0.85 & 1.87 & 1.45 & 0.81 & 1.79 \\
    $10^4$  & 1.57 & 1.12 & 1.40 & 1.41 & 0.85 & 1.65 \\
    $10^5$  & 1.57 & 1.27 & 1.24 & 1.38 & 1.01 & 1.37 \\
    $10^6$  & 1.63 & 1.63 & 1.00 & 1.37 & 1.27 & 1.08 \\
    $10^7$  & 1.47 & 1.53 & 0.96 & 1.28 & 1.10 & 1.17 \\ 
  \end{tabular}
 \end{ruledtabular} 
  \end{table}
In Table~\ref{tab_flux} presented  are neutrino flux ratios (averaged over zenith angles) calculated with usage of three hadronic models QGSJET-II, SIBYLL and KM: columns marked as $1, 2, 3$  present comparative ($\nu_\mu+\bar\nu_\mu$) fluxes,
$\phi_{\nu_{\mu}}^{(\rm SIBYLL)}/\phi_{\nu_\mu}^{(\rm QGSJET\,II)}$,  
 $\phi_{\nu_{\mu}}^{(\rm KM)}/\phi_{\nu_\mu}^{(\rm QGSJET\,II)}$,  and   
 $\phi_{\nu_{\mu}}^{(\rm SIBYLL)}/\phi_{\nu_\mu}^{(\rm KM)}$  respectively. 
Columns $4, 5, 6$ give  the same for the conventional $\nu_e +\bar\nu_e$ flux. 
All computations are performed for ZS and HGm cosmic ray spectra.  
One can see that QGSJET-II and SIBYLL 2.1 lead to apparent difference in the muon neutrino flux, as well as in the case of SIBYLL as compared to KM. However with the energy rise the flux difference between  SIBYLL and KM predictions diminishes, so that at energies above $100$ TeV these fluxes are in close agreement.
 Quite contrary, the QGSJET-II and KM flux difference becomes notable just above $100$ TeV. 

In Figs.~\ref{qgs_2pcr}, \ref{numu_sib}, the  conventional $\nu_\mu+\bar\nu_\mu$  fluxes averaged over zenith angles  in the range $96^\circ-180^\circ$ (the  upward neutrinos, $\cos\theta \alt -0.1$), calculated with use of ZS and BK spectra, are compared with Frejus~\cite{frejus}, AMANDA-II~\cite{amanda10}, and IceCube40~\cite{icecube11} measurement data obtained with the 40-string configuration of the IceCube detector. 

To illustrate the influence of the ``knee'' in  the cosmic ray spectrum, in Fig.~\ref{qgs_2pcr} we plot also the conventional ($\nu_\mu+\bar\nu_\mu$)  spectrum  computed  with usage of the cosmic ray parametrization by Gaisser, Honda, Lipari and Stanev (GH)~\cite{gh02}. The right panel of Fig.~\ref{qgs_2pcr} shows an influence  of the hadronic models, QGSJET vs SIBYLL (see also Table~\ref{tab_flux}).  
The difference in the neutrino flux predictions resulted from cosmic ray spectra becomes apparent at high neutrino energies: the flux obtained for GH spectrum  at $600$ TeV is  nearly twice as large as that for ZS  spectrum for the same hadronic model. Close to $1$ PeV  this discrepancy increases to the factor five. More results concerning the muon neutrino calculations compared to the AMANDA and IceCube40 measurements were presented in Refs.~\cite{SPS11,PSS12, SOS13}.  

The prompt neutrino flux was calculated using the quark-gluon string model (QGSM) by Kaidalov \& Piskunova~\cite{KP, bnsz89} as well as the recombination quark-parton model (RQPM)~\cite{bnsz89}. 
In Fig.~\ref{numu_sib} the prompt neutrino calculations were performed with the parametrization of cosmic ray spectrum by Nikolsky, Stamenov and Ushev (NSU)~\cite{NSU}, therefore they can serve here as upper limits for the prompt neutrino flux due to QGSM  or  RQPM. 

Addition of the  QGSM prompt contribution evidently improves the agreement  of the calculation example ZS+QGSJET with the IceCube40 measurement data~\cite{icecube11} above $100$ TeV. 
The prompt neutrino flux due to QGSM and RQPM  can be approximated at energies $5$ TeV $\leq E \leq 5$ PeV by the expressions 
 \begin{equation}\label{pms1}
\phi_{\nu}^{\rm (qgsm)}(E)
= A(E/E_1)^{-3.01}[1+(E/E_1)^{-2.01}]^{-0.165},
 \end{equation} 
\begin{equation}\label{pms2}
\phi_{\nu}^{\rm (rqpm)}(E)
 = B(E/E_1)^{-2.96}[1+(E/E_1)^{-1.96}]^{-0.157},
\end{equation}
 where $A=1.19\cdot 10^{-18}$\,, $B=4.65\cdot 10^{-18}\,\mathrm{(GeV\,cm^{2}\,s\,sr)^{-1}}$, $E_1=100$ TeV.

\begin{figure}[!b]  
\includegraphics[width=85 mm]{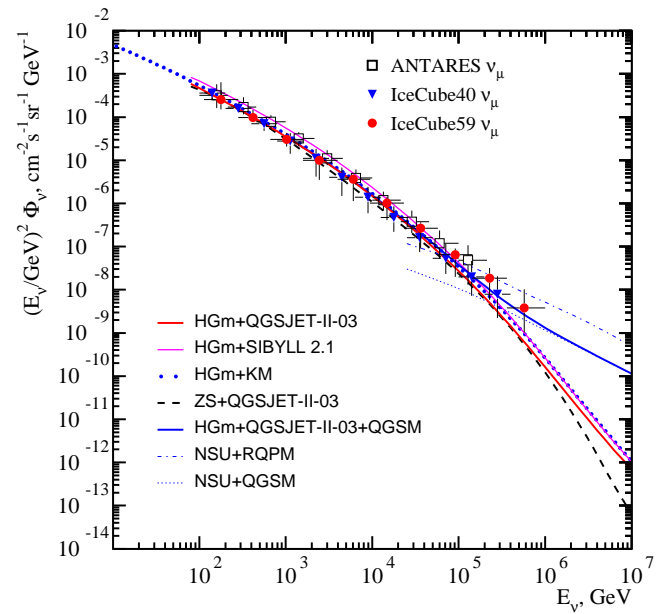} 
 \caption{\label{antares} The atmospheric $\nu_\mu+\bar\nu_\mu$  flux calculations  versus the experiment.   Conventional flux:    HGm+QGSJET (red solid line), HGm+SIBYLL (thin magenta), HGm+KM (bold dots), ZS+QGSJET (dashed). Prompt neutrinos (separately): QGSM (short dash curve),   RQPM (dash-dot). Total neutrino flux (sum of the prompt flux and conventional one):  HGm+QGSJET+QGSM (blue solid line). The measurement data (symbols): IceCube40~\cite{icecube11}, IceCube59~\cite{IC59_numu} and  ANTARES~\cite{antares13a}.}  
\end{figure}  

Figure~\ref{antares} shows calculations of the atmospheric muon neutrino spectrum   
in comparison with the measurement data obtained in recent experiments, IceCube40~\cite{icecube11},  ANTARES ~\cite{antares13a} and  IceCube59~\cite{IC59_numu}. In Ref.~\cite{IC59_numu} results were presented of the novel reconstruction of the muon neutrino spectrum with use of the data obtained with the 59-string IceCube configuration.  The calculations of the conventional flux were performed for a set of models:  HGm+QGSJET-II (red solid line), HGm+SIBYLL (magenta), HGm+KM (blue bold dots), ZS+QGSJET-II (dashed line). The  QGSM and RQPM  prompt neutrino fluxes obtained~\cite{bnsz89} with NSU spectrum are shown separately (short dash and dash-dot lines, respectively). The total atmospheric ($\nu_\mu+\bar\nu_\mu$) spectrum was calculated with the  model  HGm+QGSJET+QGSM (blue solid line).

The conventional flux due to the HGm+QGSJET is similar to that of ZS+QGSJET up to $1$  PeV. 
The difference of the neutrino flux predictions originated  from the primary cosmic ray spectra becomes apparent above $1$ PeV: the flux obtained with QGSJET-II for  ZS spectrum at $2$ PeV is less by a third  of the flux for HGm spectrum. 

The HGm+KM calculation represents, in fact, a kind of an interpolating flux (between QGSJET-II-03 flux and SIBYLL 2.1 one) because the HGm+KM prediction is in close agreement with HGm+QGSJET one at lower energies and agrees well with HGm+SIBYLL above $100$ TeV; the blue bold dot curve and thin magenta one merge into one curve at $E>200$ TeV (Fig.~\ref{antares}).

Being in a close agreement with the IceCube59 measurement data in the energy range from $180$ GeV to $36$ TeV,  the HGm+QGSJET model (only conventional flux) leads to a systematic deviation from the experimental data  of the IceCube59 muon neutrino flux, thus leaving  the window for the prompt neutrinos. 
The HGm+QGSJET+QGSM calculation of the total atmospheric  $\nu_\mu$  flux (blue solid line) may be considered as  the  preferred model, because it describes well the IceCube40 data and gives slightly lowered flux at highest energies reached in IcecCube59 experiment. Similar result gives also the  HGm+KM+QGSM model. 
 
Note that possible difference of the calculated conventional $\nu_\mu +\bar\nu_{\mu}$ spectrum, resulting from averaging over the zenith angle range $96^\circ-180^\circ$ (corresponds to the IceCube40 
zenith angle cut, $\theta>97^\circ$~\cite{icecube11}), which differs from the extended range ($\theta>88^\circ$) in the IceCube59 analysis~\cite{IC59_numu}, really  does not exceed $13$\%.

The atmospheric muon neutrino spectrum reconstructed with the IcuCube59 data reaches energies above $500$ TeV where expected is noticeable admixture of the prompt neutrinos and/or the astrophysical.  Because the Icecube59 data lead to higher neutrino spectrum (above $10$ TeV) as compared to the  IcuCube40 one, there is no compelling evidence against the QGSM prompt neutrino flux prediction, since   
HGm+QGSJET-II+QGSM gives the total atmospheric $\nu_{\mu}+\bar\nu_{\mu}$ flux not exceeding the IcecCube59 data at the highest energy ($E_\nu=575$ TeV) if the IceCube astrophysical flux was zero. 

On the contrary, if  the prompt neutrino flux is the negligible component,
then adding the best-fit astrophysical flux (Eq.~(\ref{IC_fit1}), Sect.~\ref{sec:nue}) to the lowest conventional flux prediction (ZS+QGSJET),  leads evidently to the higher flux at $575$ TeV, $\sim 1.0\cdot 10^{-8}$ (GeV\,cm$^{-2}$\,s$^{-1}$\,sr$^{-1}$), unlike the value $0.38\cdot 10^{-8}$ in the IceCube59 experiment~\cite{IC59_numu}.  
Conceivably this means that the hypothesis of the flavor equipartion ($\nu_e:\nu_\mu:\nu_\tau=1:1:1$)  of the IceCube  astrophysical neutrino flux needs a revision (see Refs.~\cite{aaron14, watanabe14}). 
\begin{table}[!b]
\caption{\label{conv_param}Parameters for the conventional neutrino spectra (Eq.~(\ref{canu})).}
\begin{ruledtabular}
\begin{tabular}{ccccc} 
  flavor   & $C_{\nu}$ & $\gamma_{0}$ & $\gamma_{1}$ & $\gamma_{2}$ \\  \hline 
 $\nu_\mu +\bar\nu_\mu$ & $4.896\cdot10^{-3}$ & $2.198$ & $1.648\cdot 10^{-1}$ & $1.46\cdot10^{-3}$ \cr
 $\nu_e +\bar\nu_e$     & $6.053\cdot10^{-3}$ & $2.918$ & $4.899\cdot 10^{-2}$ & $7.25\cdot10^{-3}$  \cr 
 \end{tabular}  
\end{ruledtabular} 
\end{table}

We can describe the conventional zenith-angle averaged $\nu_\mu+\bar\nu_\mu$ flux (due to the prediction of the  preferred model, HGm+QGSJET-II)  by the approximation, valid for $10^2 - 10^7$ GeV  with errors not exceed $12$\% (at lower energies) (in units of  $\mathrm{cm^{-2}s^{-1}sr^{-1}GeV^{-1}}$):  
\begin{equation}\label{numu_approx}
\lg[E^2_{\nu}\phi^{\pi,\,K}_{\nu_\mu}(E_{\nu})]
=-(2.31+0.198y+0.165y^2+0.00146y^3),
 \end{equation}
where $y=\lg(E_{\nu}/1\,\rm GeV).$ 
 Here we also present the approximation formula describing our calculation (Sec.~\ref{sec:nue})  for the atmospheric conventional $\nu_e+\bar\nu_e$ spectrum (the same model, HGm+QGSJET-II-03): 
 \begin{equation}\label{nue_approx}
\lg[E^2_{\nu}\phi^{\pi,\,K}_{\nu_e}(E_{\nu})]
=-(2.218+0.918y+0.04899y^2+0.00725y^3),
\end{equation}
\noindent
 Fluxes of neutrinos  Eqs.~(\ref{numu_approx}), (\ref{nue_approx})  can be rewritten  also in the form:
 \begin{equation}\label{canu}
\phi_{\nu}^{\pi, K}(E_\nu)
 = C_{\nu}\left(\frac{E_{\nu}}{1\,\rm GeV}\right)^{-(\gamma_{0}+\gamma_{1}y+
\gamma_{2}y^{2})}.
\end{equation}
Two sets of the parameters to Eq.~(\ref{canu}) are given in Table~\ref{conv_param}.

\begin{table}[!b]
\caption{\label{tab_3}Atmospheric neutrino flux in the energy range  $(0.4-1)$ PeV  and  upper limit for diffuse $\nu_\mu+\bar\nu_\mu$ flux obtained with neutrino telescopes.} 
\begin{ruledtabular}   
   \begin{tabular}{ll}  
     Model & $E_\nu^2\phi_\nu$, GeV\,(cm$^{2}$\,s\,sr)$^{-1}$  \\ \hline
  conventional $\nu_\mu+\bar\nu_\mu$: &  $E_\nu=400$ TeV - $1$ PeV  \\ 
   averaged over angles -- &    \\ 
  ZS+SIBYLL 2.1   &  $ (2.21 - 0.214)\times 10^{-9}$ \\ 
  ZS+QGSJET-II &  $(1.32 - 0.149)\times 10^{-9}$ \\  
  BK+QGSJET~II &  $(1.09 - 0.097)\times 10^{-9}$ \\ 
  HGm+QGSJET~II          &  $(1.45 - 0.163)\times 10^{-9}$ \\ \hline 
        selected zenith angles:    &   $400$ TeV - $1$ PeV \\
  HGm+QGSJET-II, $\cos\theta=0.5$  &  $(0.97-0.109)\times 10^{-9}$   \\ 
  HGm+QGSJET-II, $\cos\theta=0.3$  &  $(1.56-0.176)\times 10^{-9}$   \\ 
  HGm+QGSJET-II, $\cos\theta=0.1$  &  $(3.40-0.384)\times 10^{-9}$   \\ 
     \hline  
  prompt $\nu_\mu+\bar\nu_\mu$ :  &   $400$ TeV - $1$ PeV  \\  
  NSU+QGSM   & $(2.90 - 1.16) \times 10^{-9}$  \\
  ZS+QGSM    & $(2.23 - 0.54) \times 10^{-9}$  \\  
 $174E_{N}^{-3}+$ DM~\cite{DM} & $(1.87 - 0.85) \times 10^{-9}$  \\   \hline 
  conv. + prompt $\nu_\mu+\bar\nu_\mu$:   & $400$ TeV - $1$ PeV \\
  conv.(averaged)+ QGSM & $(4.35-1.32)\times 10^{-9}$  \\ 
  conv.(averaged)+ DM   & $(3.32 - 1.01) \times 10^{-9}$  \\
  conv.($\cos\theta=0.1$)+QGSM & $(6.30-1.54) \times 10^{-9}$  \\ \hline 
  diffuse $\nu_\mu+\bar\nu_\mu$:  & $34.5$ TeV –- $36.6$ PeV \\  
  IC59 best fit~\cite{ice59_lim}  & $0.25 \times 10^{-8}$    \\ 
  IC59 limit~\cite{ice59_lim}     & $1.44 \times 10^{-8}$     \\
                                  & $45$ TeV –- $10$ PeV       \\ 
  ANTARES limit~\cite{antares13b} & $4.8 \times 10^{-8}$       \\  
   \end{tabular}  
\end{ruledtabular}
  \end{table}   

The atmospheric muon neutrino fluxes,  calculated in the  energy range $0.4-1$ PeV, are presented also in Table~\ref{tab_3} along with upper limits on the diffuse  flux of  astrophysical muon neutrinos  obtained   in  the   ANTARES~\cite{antares13b} and  IceCube59 ~\cite{ice59_lim} experiments.
The total atmospheric muon neutrino flux (sum of the conventional flux and prompt one) marked in Table~\ref{tab_3} as ``conv.~(averaged)+ QGSM'' is presented by the preferred model HGm+QGSJET-II + QGSM.
Note the prompt neutrino flux obtained with the dipole model (DM)~\cite{DM} is close to to the QGSM prediction~\cite{bnsz89} above  $1$ PeV (about $30$\% of the disagreement). 
More intent inspection of the predicted atmospheric neutrino fluxes, both conventional and prompt,  shows that distinctions between the DM and QGSM prompt muon neutrino flux predictions are hardly observable at the present level of experimental errors, if the knee of the cosmic ray spectrum is thoroughly taken into account.

\section{Electron neutrino flux and the neutrino flavor ratio \label{sec:nue}} 

The sources of the conventional $\nu_e$ are three-particle decays of muons $\mu_{e3}$,  and kaons $K^{\pm}_{e3}$ , $K^{0}_{e3}$   with the branching ratio  $5.07\%$ and $40.5\%$ respectively. The latter is dominant source of electron neutrinos at energy below $10$ TeV. The semileptonic decays  of short-lived $K^{0}_{S}$ also contribute though branching ratio of the decay is small ($0.07\%$)~\cite{footnoteK0S}.
This decay gives a considerable contribution to the atmospheric  $\nu_e$ flux  at high energies, reaching  $36$\% at $E_\nu=500$ TeV for zenith angle $\theta=0^\circ$  (HGm+QGSJET-II model). 
Close to vertical the ($\nu_e+\bar\nu_e$) flux from the $K^{0}_{S}$ decay becomes nearly equal to that from $K^{0}_{L}$ one at $E_\nu\approx 1$ PeV.
\begin{figure}[!b]
\includegraphics[width=85 mm]{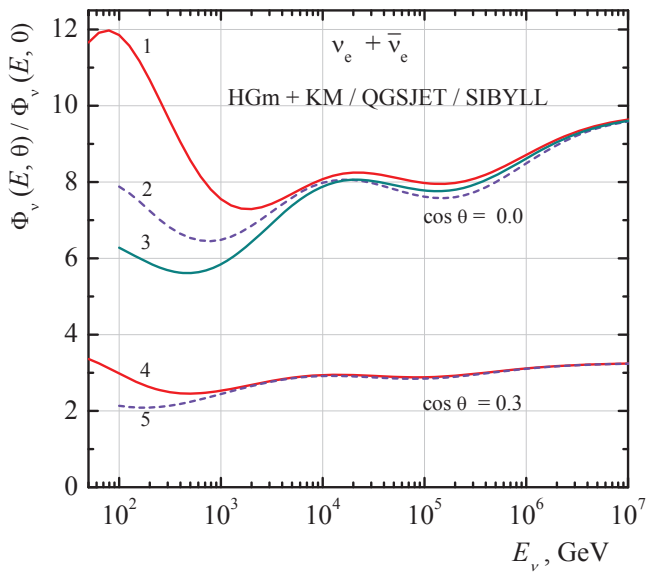} 
\caption{\label{angle_nue} Zenith-angle enhancement  of the atmospheric ($\nu_e+\bar\nu_e$) flux due to energy alignment of the kaon sources. Curves represent the flux ratio, $\phi_{\nu_e}(\theta)/\phi_{\nu_e}(0^\circ)$, calculated with the HGm primary spectrum for 
$\theta=72.5^\circ$,   $90^\circ$:  $1$ --  KM hadronic model  ($90^\circ$), $2$ --  QGSJET-II-03 ($90^\circ$), $3$ --  SIBYLL 2.1 ($90^\circ$), $4$ --  KM ($72.5^\circ$), $5$ --  QGSJET-II-03  ($72.5^\circ$).}    
\end{figure}

Figure~\ref{angle_nue} shows  zenith-angle-dependent $\nu_e$ flux ratio, $\phi_\nu(E,\theta)/\phi_\nu(E,0^\circ)$, calculated for $\theta=90^\circ$ and $72.5^\circ$) with usage of primary spectrum HGm  and  hadronic interaction models KM, QGSJET-II-03 and SIBYLL 2.1. The  curves illustrate partial contributions of kaon sources varying with the energy in accordance with  
the scale -- the  critical energy depending on the decay mode and zenith angle: 
$\epsilon^{cr}_{K^0_{L}} (\theta=0^\circ)=210$ GeV, $\epsilon^{cr}_{K^\pm_{\ell3}} (\theta=0^\circ)=890$ GeV, $\epsilon^{cr}_{K^0_{S}} (\theta=0^\circ)=120$ TeV (the critical energy for  the horizontal is one order of magnitude larger). The ``wave'' of zenith-angle enhancement of the atmospheric ($\nu_e+\bar\nu_e$) flux makes apparent the successive ``switching-on'' of the kaon sources.

The approximation formula, Eq.~(\ref{canu}), describing the calculated zenith-angle averaged energy spectrum  of the atmospheric $\nu_e+\bar\nu_e$  flux (HGm+QGSJET-II),  was given in Sec.~\ref{sec:numu}.   

Recently the IceCube published results~\cite{ice79_nue} of  the first measurement of the atmospheric electron neutrino spectrum in the energy range $80$ GeV - $6$ TeV obtained with the 79-string IceCube configuration including DeepCore. These measurement data make possible an evaluation of the neutrino flavor ratio and comparison it with  predictions. 

\begin{figure}[!t]
\includegraphics[width=85 mm]{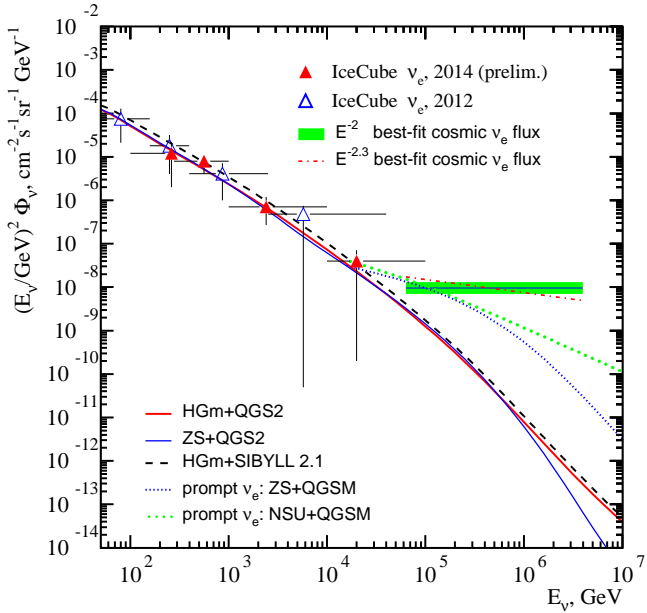} 
\caption{\label{IC_PeVnu} Atmospheric ($\nu_e+\bar\nu_e$) spectrum and the diffuse flux of cosmic neutrinos observed in IceCube experiment~\cite{3years}. The IceCube experimental data:  atmospheric neutrinos  \cite{ice79_nue} (triangles) and \cite{ic2014_cern} (filled triangles); the band width reflects the statistical uncertainty of the IceCube best fit for astrophysical neutrino flux~\cite{3years}, Eq.~(\ref{IC_fit1});  the red dash-dot line corresponds to the IceCube best-fit power law for the astrophysical neutrino spectrum~\cite{3years}, Eq.~(\ref{IC_fit2}). Curves: predicted fluxes of the atmospheric conventional and prompt neutrinos.}  
\end{figure}

In Fig.~\ref{IC_PeVnu} we compare the atmospheric  ($\nu_e+\bar\nu_e$) flux  calculated using QGSJET~II-03 and SIBYLL 2.1  with  IceCube  measurement data~\cite{ice79_nue} (open triangles) and, besides,  with recent  IceCube preliminary data for the atmospheric electron neutrinos ~\cite{ic2014_cern} (red fill triangles). 

The diffuse flux of cosmic neutrinos based on 37 events observed in IceCube  experiment~\cite{science342,3years} is presented in Fig.~\ref{IC_PeVnu} (the green band and red dash-dot line) with use of the IceCube best fits per flavor, $\phi_\nu \sim E^{-2}$ and $\phi_\nu \sim E^{-2.3}$, borrowed from Ref.~\cite{3years}: 
 \begin{equation}\label{IC_fit1}
E^2\phi_\nu=(0.95\pm 0.3)\cdot 10^{-8} \,  {\rm GeV} {\rm cm}^{-2}{\rm s}^{-1}{\rm sr}^{-1};  	
\end{equation}
\begin{equation}\label{IC_fit2}
  E^2\phi_\nu=1.5\cdot10^{-8}(E/100~{\rm TeV})^{-0.3}\,{\rm GeV}/({\rm cm}^{2} {\rm s}\, {\rm sr}).  
 \end{equation}
The equal flavor composition ($\nu_e:\nu_\mu:\nu_\tau=1:1:1$) and  zero prompt neutrino flux 
were supposed to derive the fits  which are valid for deposited energies of neutrino events in the range  $60$ TeV$<E<3$ PeV.
The flavor composition of the IceCube astrophysical high-energy neutrino flux  is discussed in detail in Refs.~\cite{aaron14, watanabe14, fargion13, fu14, xun14, chen14}.
%
%
\begin{figure}[!t]  
\centering \hskip 0.2 cm
\includegraphics[width=85 mm] {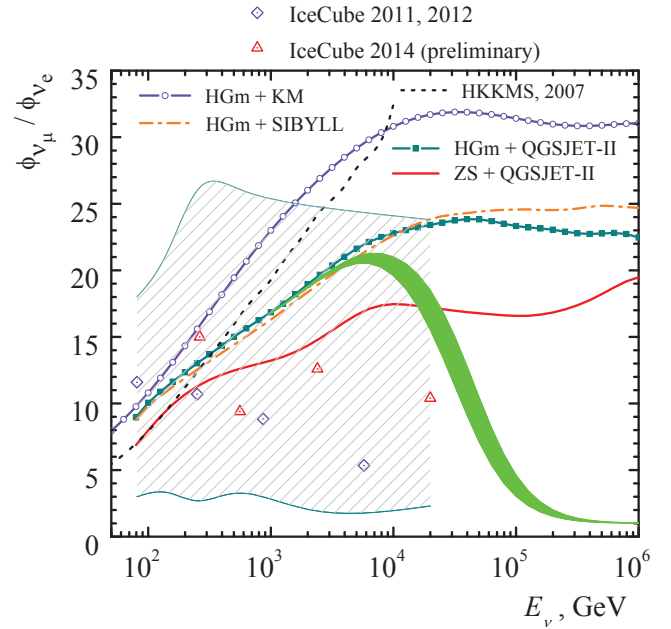} 
\caption{\label{flavor-1} 
The atmospheric neutrino flavor ratio $R_{\nu_\mu/\nu_e}$ calculated for several models in comparison with results derived from IceCube experiments~\cite{icecube11, ice79_nue, ic2014_cern}.
Symbols correspond to $R_{\nu_\mu/\nu_e}$ reconstructed from the IceCube measurement data,  hatched area images  these data uncertainties.  The filled green area represents $R_{\nu_\mu/\nu_e}$ obtained for sum of the conventional neutrino flux prediction (HGm+QGSJET) and the best-fit astrophysical neutrino flux (Eq.~(\ref{IC_fit1})).} 
\end{figure}

The curves in Fig.~\ref{IC_PeVnu} correspond to atmospheric electron neutrino flux, calculated separately for the conventional neutrinos (HGm/ZS+QGSJET/SIBYLL) and prompt ones. 
 The QGSM prompt neutrino flux is represented in Fig.~\ref{IC_PeVnu} with two curves:  green bold dots correspond to the original QGSM flux~\cite{bnsz89}, Eq.~(\ref{pms1}), and blue short dash line corresponds to that rescaled with the ZS cosmic ray spectrum.
 It is clearly seen that the astrophysical $\nu_e$ flux and the atmospheric prompt one are competing contributions into detector events at energies above $30-50$ TeV. 

Since IceCube has measured energy spectra both of muon and electron neutrinos, one can try to construct the neutrino flavor ratio $R_{\nu_\mu/\nu_e}=\phi_{\nu_\mu+\bar\nu_\mu}/\phi_{\nu_e+\bar\nu_e}$ and  check for agreement of the calculation  with experimental data. 	
 The flavor ratio, being responsive to  changes of the electron neutrino flux, allows to reveal an admixture of neutrinos from astrophysical sources or an addition of the atmospheric prompt neutrinos.

The difference in predictions of atmospheric neutrino flux ratio related to choice of hadronic models  is seen in Fig.~\ref{flavor-1}: curves display the scale of difference of the conventional neutrino spectra, calculated with models QGSJET-II, SIBYLL 2.1 and KM for HGm and ZS parametrizations of cosmic-ray spectra (HGm+KM/QGSJET/SYBILL, ZS+QGSJET). 
The  dashed curve (HKKMS, 2007) in Fig.~\ref{flavor-1} shows the Monte Carlo calculation by 
Honda et al.~\cite{HKKM07} made with usage of the hadronic model DPMJET-III~\cite{mc_code00, dpmjet07},
the top curve corresponds to  $R_{\nu_\mu/\nu_e}$ calculated with KM hadronic model.

This figure displays  visible difference of DPMJET-III and KM predictions on the one hand,  QGSJET II and  SIBYLL  on the other, which is partly  attributable to the difference of hadronic models in kaon yield
(Table~\ref{tab_z}). 
Relative proximity of  $z$-factors for KM  and DPMJET-III models leads to similar behavior of the neutrino flavor ratio, $R_{\nu_\mu/\nu_e}$ (Fig.~\ref{flavor-1}). 
However the  $R_{\nu_\mu/\nu_e}$ dissimilarity between SIBYLL and QGSJET-II  is not so large  as would be expected from the large difference in the kaon production.
On the contrary, the same model (QGSJET-II) leads also to visible  $R_{\nu_\mu/\nu_e}$ distinction arising  from the cosmic ray spectrum (ZS vs. HGm): HGm+QGSJET-II (the curve with dark cyan small squares)  and   ZS+QGSJET-II (solid  red  curve). 

Relative  kaon excess due to KM,  DPMJET-III  and SIBYLL makes some kind  ``hierarchy'' of  the plateau in the energy dependence of  $R_{\nu_\mu/\nu_e}$  (still corrected for different pion yield):
$$ R_{\nu_\mu/\nu_e}^{\text{(QGSJET)}}< R_{\nu_\mu/\nu_e}^{\text{(SIBYLL)}}<  R_{\nu_\mu/\nu_e}^{\text{(DPMJET)}} < R_{\nu_\mu/\nu_e}^{\text{(KM)}}.$$
 
Figure~\ref{flavor-1} shows also a reconstruction of $R_{\nu_\mu/\nu_e}$  derived from the atmospheric neutrino spectra, measured by IceCube~\cite{icecube11, ice79_nue}.
Diamonds denote $R_{\nu_\mu/\nu_e}$ values reconstructed with usage of the IceCube data~\cite{ice79_nue}
and~\cite{icecube11}, triangles correspond to the latest preliminary IceCube data for the atmospheric electron neutrino spectrum~\cite{ic2014_cern}.   The hatched area displays the  $R_{\nu_\mu/\nu_e} $ statistical uncertainties  of the measurement data:
 $$\delta R_{\nu_\mu/\nu_e} = R_{\nu_\mu/\nu_e} \sqrt{(\delta \phi_{\nu_\mu} /\phi_{\nu_\mu})^2+(\delta \phi_{\nu_e}/\phi_{\nu_e})^2} .$$
 
To derive  $\nu_\mu$ and $\nu_e$ flux values at equal energies, a local interpolation of the experimental data is required.  This problem was solved by making the interpolating function (a local fit) for the $\nu_\mu$ spectrum at energies $100$ GeV - $20$ TeV supported by IceCube40 $\nu_\mu$ measurement data~\cite{icecube11} in the range  $141$ GeV -  $35.5$ TeV.  As the denominator in the $R_{\nu_\mu/\nu_e}(E)$ ratio taken were namely experimental values  of the IceCube $\nu_e$ flux from Ref.~\cite{ice79_nue}, and separately from Ref.~\cite{ic2014_cern}.  The robustness of the procedure was tested by use, on the contrary, of the $\nu_\mu$ flux measurement data and a fit for the $\nu_e$ spectrum. 

In the range $10-100$ TeV our calculations with QGSJET-II and SIBYLL give $R_{\nu_\mu/\nu_e}$ values  $17-25$  depending on the cosmic ray spectra (compare ZS+QGSJET-II and HGm+QGSJET-II/SIBYLL, Fig.~\ref{flavor-1}), while  the  Monte Carlo computation (using the CORSIKA) for the same models~\cite{fedyn12} gives larger values, $25-32$.

The hadronic model by Kimel and Mokhov leads to the neutrino flavor ratio $R_{\nu_\mu/\nu_e}$
rather close to result of Ref.~\cite{HKKM07} for DPMJET-III~\cite{mc_code00, dpmjet07} in the range $100$ GeV - $10$ TeV.  In Fig.~\ref{flavor-1} we see, however, that values of the  flavor ratio, like those due to the HGm+KM calculation, $R_{\nu_\mu/\nu_e}\approx 30 $ at $10-30$ TeV, exceed noticeably not only  $R_{\nu_\mu/\nu_e}$  midpoints (symbols in Fig.~\ref{flavor-1}), reconstructed from IceCube data, but also  the hatched area. 

The neutrino flavor ratio calculated with QGSJET-II and SIBYLL 2.1 models agrees on the whole (within statistical and systematic uncertainties) with  that reconstructed from IcecCube data. However the IceCube $R_{\nu_\mu/\nu_e}$ midpoints (symbols in Fig.~\ref{flavor-1}) do not reveal a trend toward rising as was expected for the conventional neutrino flux in the energy range $100$ GeV -- $30$ TeV. This behavior possibly indicates  that the atmospheric electron neutrino flux, measured in the IceCube experiment, contains an admixture of the astrophysical neutrinos yet in the range $10-30$ TeV.

The  curved band in Fig.~\ref{flavor-1} is obtained as sum of the calculated conventional neutrino flux 
(HGm+QGSJET-II, zero prompt neutrino flux) and the astrophysical one in accordance with IceCube best fit~\cite{3years} (Eq.~(\ref{IC_fit1})). 
To compute $R_{\nu_\mu/\nu_e}(E_\nu)$ in the range below $60$ TeV, the IceCube best fit was extrapolated to $10$ TeV. No prompt neutrinos were taking into account. The width of the band reflects the IceCube astrophysical flux uncertainty, $\pm 0.3 \cdot 10^{-8}\, {\rm GeV} {\rm cm}^{-2}{\rm s}^{-1}{\rm sr}^{-1}$. The extrapolation of the best-fit astrophysical neutrino flux to the energy range below $60$ TeV does not contradict $R_{\nu_\mu/\nu_e}$ reconstructed from the IceCube data for atmospheric neutrinos.

In case of zero prompt component, the IceCube best-fit astrophysical neutrino flux  dominates over atmospheric ($\nu_e+\bar\nu_e$) flux at energies above $50$ TeV. Close to $50$ TeV, the predicted (HGm+QGSJET) atmospheric conventional $\nu_e$ flux (scaled by $E^2$) is about $0.46\cdot 10^{-8}  {\rm GeV} {\rm cm}^{-2}{\rm s}^{-1}{\rm sr}^{-1}$, which is less than, by half, of the best-fit astrophysical flux. 
Alternative hypothesis, allowing for the prompt neutrino  component, leads to similar depression of $R_{\nu_\mu/\nu_e}$ and also might be accepted.  However, in this case the prompt neutrino flux should  be rather large one, like the QGSM prediction.

\section{Summary}  

The problem of the atmospheric neutrino background  became really important after observation with the IceCube detector of events induced by very high-energy neutrinos of extraterrestrial origin~\cite{pevnu2013,science342,3years}.
More  precise  calculations of the high-energy neutrino spectrum  are  required  since it is possible that astrophysical  neutrinos are entangled with the atmospheric neutrinos arising from decays of $\pi$, $K$ mesons (conventional neutrinos) and  decays of charmed particles (prompt neutrinos) which are produced
in collisions of cosmic ray particles with the  Earth's atmosphere.
  
Presented in this work  are the results of the calculations of the high-energy atmospheric neutrino fluxes  performed for hadronic interaction models  QGSJET-II-03, SIBYLL 2.1 and Kimel \& Mokhov, taking into consideration the ``knee'' of the cosmic-ray  spectrum.  
The calculation shows rather weak dependence on the cosmic ray spectrum in the energy range $10^2-10^5$ GeV. However, the picture appears to be less steady because of sizable difference of the hadronic models predictions. As can be seen in the example of the models QGSJET~II-03 and SIBYLL 2.1, the major factor in the discrepancy in conventional neutrino fluxes is the kaon production in nucleon-nucleus collisions.
Really,  cosmic-ray  physicists feel necessity of comprehensive analysis  of the actual features of the high-energy hadronic models under discussion, QGSJET II-03 (04), EPOS-LHC, SIBYLL 2.1, DPMJET-III, especially  concerning details of the kaon and  charmed particle production in $N A$, $\pi A$ collisions.

Above $100$ TeV calculated spectra of muon neutrinos  display an apparent dependence on the spectrum and composition of primary cosmic rays related to  the ``knee'' range. 
Also in this region, uncertainties appear due to production cross sections and decays of charmed particles which imprint on the prompt neutrino flux. 

All calculations  are  compared with the atmospheric neutrino  measurements  by  Frejus, AMANDA, IceCube and ANTARES.  New reconstruction of the $\nu_\mu$  spectrum,  performed by IceCube Collaboration~\cite{IC59_numu}, seemingly does not map out the QGSM prompt neutrino flux prediction.
Being in a close agreement with the IceCube measurement data in the energy range from $140$ GeV to $100$ TeV,  the HGm+QGSJET model leads to the systematic deviation from experimental data, especially those for the IceCube59 muon neutrino spectrum in energy the range $100-500$ TeV.  Thus, the IceCube59 data leave a window for the  QGSM prompt neutrino component: the comparison of the calculation with IceCube measurement data on atmospheric $\nu_\mu$ and $\nu_e$ fluxes makes it possible to consider the HGm+QGSJET-II-03+QGSM model as the preferable one.

The approximation formula describing the HGm+QGSJET-II-03 predictions of the  atmospheric conventional $\nu_e+\bar\nu_e$  and ($\nu_\mu+\bar\nu_\mu$) energy  spectra, averaged over  zenith angles, is given by Eq.~(\ref{canu}). An analytic description of $\nu_\mu+\bar\nu_\mu$ and $\nu_e+\bar\nu_e$ energy spectra HGm+QGSJET-II-03 can be used, in principle, as one more tool to test data of neutrino event reconstruction in neutrino telescopes. 

Authors of the IceCube59 analysis~\cite{IC59_numu} avoid definite conclusions concerning 
the prompt neutrino contribution or the neutrinos of a cosmic origin because of large systematic uncertainties at the highest energies. Nevertheless, the $\nu_\mu$ flux mean values illustrate the state of the problem. Even if the prompt neutrino flux is zero, sum of the best-fit astrophysical flux and  the calculated  atmospheric conventional one, gives the evidently higher flux as compared to the IceCube59 data (the last  bin around $E_\nu\approx 575$ TeV). 
This circumstance gives rise to doubts regarding the equal flavor composition of the IceCucbe astrophysical flux, $\nu_e:\nu_\mu:\nu_\tau=1:1:1$.

The IceCube best-fit astrophysical neutrino flux  dominates over atmospheric ($\nu_e+\bar\nu_e$) flux at energies above $60$ TeV (Fig.~\ref{IC_PeVnu}). Around $60$ TeV, the highest  predicted (HGm+SIBYLL) atmospheric conventional $\nu_e$ flux (scaled by $E^2$) is about $0.5\cdot 10^{-8}\,  {\rm GeV} {\rm cm}^{-2}{\rm s}^{-1}{\rm sr}^{-1}$, that is well below the IceCube best-fit astrophysical flux, Eq.~(\ref{IC_fit1}). 
Thus, the transition from the atmospheric electron neutrino flux to the  predominance of the astrophysical neutrinos occurs at $30-100$ TeV if the prompt neutrino component is taken into consideration. 

The Kimel \& Mokhov hadronic model (HGm+KM) and DPMJET-III~\cite{HKKM07} lead to rather large values of the atmospheric neutrino flavor ratio in the energy range $10-100$ TeV, exceeding those reconstructed from IceCube data. The similar values of  the flavor  ratio were obtained in the MC computation~\cite{fedyn12} ($R_{\nu_\mu/\nu_e}\approx 25-32$ at $E_{\nu}=10-100$ TeV), using QGSJET-II and SIBYLL 2.1 models.
There is little doubt in this case that a discordance takes place between the calculated $R_{\nu_\mu/\nu_e}$ and the median flavor ratio reconstructed from the IceCube experiment.

At the same time, the neutrino flavor ratio calculated in the present work with QGSJET-II-03 and SIBYLL 2.1 models better agrees (especially for ZS spectrum)  with  $R_{\nu_\mu/\nu_e}$ reconstructed from  the IcecCube data. Note, however,  that the IceCube midpoints (symbols in Fig.~\ref{flavor-1}) display rather low $R_{\nu_\mu/\nu_e}$  at $1-10$ TeV, which  does not reveal the  trend to increase with energy.

If the power law $E^{-2}$  is valid for the astrophysical neutrino spectrum at energies below $60$ TeV, then
 extrapolation to lower energies of the high-energy neutrino flux, observed in the IceCube experiment\cite{pevnu2013, science342, 3years},  should lead to decrease of the neutrino flavor ratio  $R_{\nu_\mu/\nu_e}$  in the energy range $10-50$ TeV.  
This extrapolation shows the consistency of the ratio $R_{\nu_\mu/\nu_e}$, calculated with model HGm+QGSJET-II-03, and that of obtained from the IceCube data.
The computation of neutrino flavor ratio hints that one more confirmation of the presence of astrophysical neutrinos might be obtained from little progress in measurement of the $\nu_e$ spectrum above $20$ TeV, because $R_{\nu_\mu/\nu_e}$, more sensitive to the electron neutrino flux, makes it possible to disclose a small fraction from astrophysical sources.

The neutrino flavor ratio $R_{\nu_\mu/\nu_e}$ extracted from IceCube data does not display a trend
toward  rising as is expected for the conventional neutrino flux in the energy range $100$ GeV -- $30$ TeV. 
The $R_{\nu_\mu/\nu_e}$ depression possibly indicates that the atmospheric electron neutrino flux measured in the IceCube experiment contains the admixture of the cosmic neutrinos even in the energy range $10-50$ TeV.%

\bigskip   

\section*{Acknowledgments}

We thank  V.~A.~Naumov for helpful discussions and comments. Authors are grateful to A.~A.~Kochanov for considerable assistance in computations. We acknowledge the support from Ministry of Education and Science of the Russian Federation (the Agreement No. 14.B25.31.0010, zadanie 3.889.2014/K) and the Russian Foundation for Basic Research, Grant No. 13-02-00214. The research was also supported in part by the Grant of President of Russian Federation No. NS-3003.2014.2.

\end{document}